\documentclass[letterpaper]{article} % DO NOT CHANGE THIS
\usepackage[preprint]{aaai2027} % DO NOT CHANGE THIS
\usepackage[hyphens]{url} % DO NOT CHANGE THIS
\usepackage{graphicx} % DO NOT CHANGE THIS
\usepackage{natbib} % DO NOT CHANGE THIS AND DO NOT ADD ANY OPTIONS TO IT
\usepackage{caption} % DO NOT CHANGE THIS AND DO NOT ADD ANY OPTIONS TO IT
\usepackage{amsmath,amssymb}
\usepackage{booktabs}
\usepackage{subcaption}
\usepackage{xcolor}
\usepackage{algorithm}
\usepackage{algpseudocode}

\graphicspath{{Figures/}}
\newcommand{\method}{C\textsuperscript{2}T-OpenMax}

\begin{document}
\title{C\textsuperscript{2}T-OpenMax: A Novel Open-Set WiFi RF Fingerprinting Method via Center Constrained Learning and Confidence-Guided Tail Modeling}
\author{
    %Authors
    % All authors must be in the same font size and format.
    %Written by AAAI Press Staff\textsuperscript{\rm 1}\thanks{With help from the AAAI Publications Committee.}\\
    %AAAI Style Contributions by Peter Patel Schneider,
    Yuanyu Zhang\textsuperscript{\rm 1},
    Junjie Yang\textsuperscript{\rm 1},
    Ji He\textsuperscript{\rm 1},
    Shuangrui Zhao\textsuperscript{\rm 1},
    Lele Zheng\textsuperscript{\rm 1},
    Yulong Shen\textsuperscript{\rm 1}
}
\affiliations{
    %Afiliations
    \textsuperscript{\rm 1}School of Computer Science and Technology, Xidian University\\
    % If you have multiple authors and multiple affiliations
    % use superscripts in text and roman font to identify them.
    % For example,

    % Sunil Issar\textsuperscript{\rm 2},
    % J. Scott Penberthy\textsuperscript{\rm 3},
    % George Ferguson\textsuperscript{\rm 4}\corresponding,
    % Hans Guesgen\textsuperscript{\rm 5}
    % Note that the comma should be placed after the superscript

    %1101 Pennsylvania Ave, NW Suite 300\\
    %Washington, DC 20004 USA\\
    % email address must be in roman text type, not monospace or sans serif
    yyuzhang@xidian.edu.cn
%
% See more examples next
}

\maketitle

\begin{abstract}
Radio frequency fingerprinting (RFF) enables device authentication from transmitter-specific hardware imperfections, but practical deployment requires cross-environment open-set recognition. Data augmentation improves environmental generalization, yet may yield dispersed, low-confidence known-class representations that distort the class statistics used by OpenMax. To address this problem, we propose \method{}, an enhanced OpenMax framework combining center-constrained learning with confidence-guided tail modeling. The former improves intra-class compactness, making class-wise representations more suitable for distance-based modeling. The latter retains only correctly classified, high-confidence logits for mean activation vector estimation and Weibull fitting, reducing bias from ambiguous boundary samples. Together, the two modules refine representation geometry and OpenMax construction while preserving augmentation benefits. Experiments on a public WiFi CSI dataset show that \method{} achieves the highest open-set accuracy in seven of eight location groups and outperforms all baselines in area under the receiver operating characteristic curve (AUROC) and open-set classification rate (OSCR) across every tested openness level. Under the largest-openness setting, it improves accuracy by 12.31\%, AUROC by 0.0887, and OSCR by 0.0856 over the augmented OpenMax baseline.
\end{abstract}

\begin{figure*}[ht]
	\centering
	\par\smallskip
	\includegraphics[width=\textwidth,trim=0 0 0 0,clip]{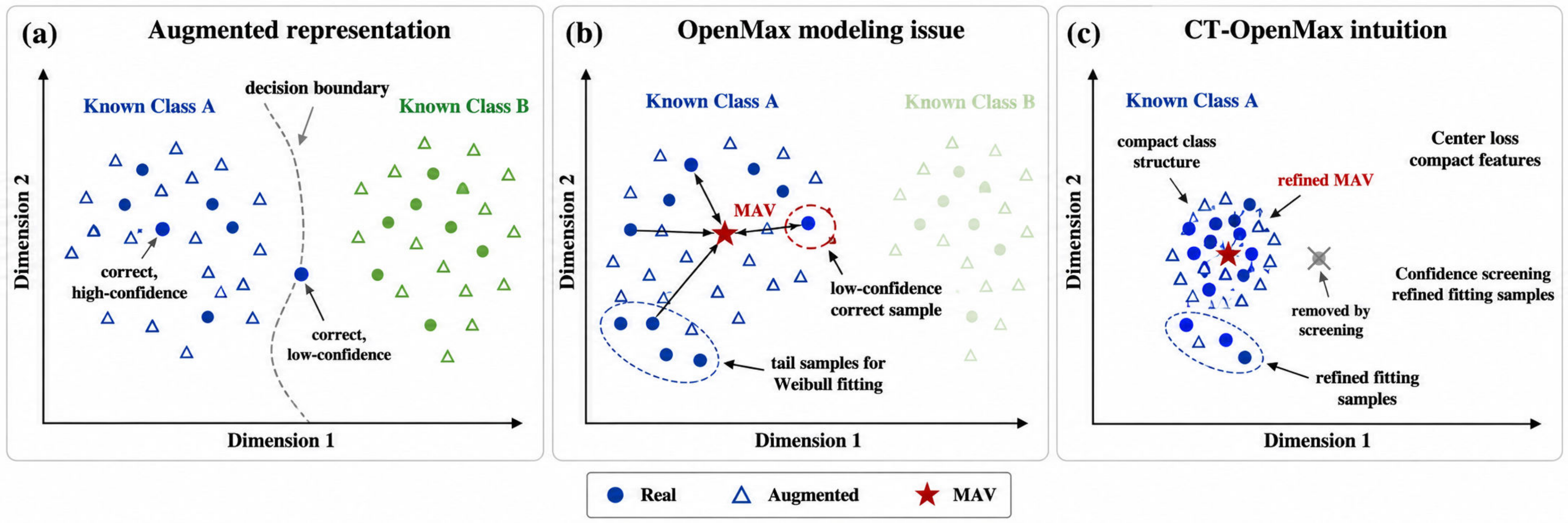}
	\par\smallskip
		\caption{Motivation of C\textsuperscript{2}T-OpenMax.
		(a) In an augmented closed-set representation, correctly classified samples
		from the same known class can differ in confidence and proximity to the
		decision boundary.
		(b) When all correctly classified samples are used for OpenMax construction,
		low-confidence boundary samples may affect both MAV estimation and the
		distance tail used for Weibull fitting.
		(c) Center-constrained learning improves intra-class compactness, while
		confidence screening refines the samples used for MAV estimation and
		Weibull tail fitting. The scatter plots are conceptual illustrations rather
		than exact activation vector distributions.}
	\label{fig:motivation}
\end{figure*}

\section{Introduction}

RFF identifies wireless devices through
physical layer features caused by device-specific hardware imperfections.
These imperfections arise from unavoidable variations in radio components
during manufacturing, making the resulting fingerprints closely related to
individual devices and difficult to reproduce exactly
~\cite{huang2017radio,xie2024radio,yan2025radio,zhang2025physical}.
For WiFi systems, channel state information (CSI) provides detailed amplitude
and phase responses, from which deep models can learn useful device
representations
~\cite{meneghello2022deepcsi,yang2023eliminating,kong2024deepcrf}.
However, most existing deep RFF methods follow a closed-set assumption,
where all devices observed during testing belong to classes included in the
training set. This assumption does not hold in practical networks, where
previously unseen devices may access the network and changes in location,
collection time, and channel conditions may alter the CSI features of the
same device. A model trained under limited environments may therefore confuse
channel changes with device differences, causing an authorized device to be
classified as another device or rejected as unknown. Meanwhile, an unknown
device may be incorrectly accepted as an authorized device. Open-set
recognition (OSR) must therefore recognize authorized devices reliably across
changing environments while effectively detecting unknown devices. Simply
assigning every sample to the known class with the highest score is no longer
sufficient for practical deployment
~\cite{zhang2022data,yu2024open,he2024open}.

Data augmentation is commonly used to improve environmental generalization
by simulating a wider range of channel conditions. However, augmentation may
also produce samples that deviate from the typical distribution of their
source devices. Although these samples may still be classified correctly,
they often exhibit low classification confidence. When used for known-class
modeling, such samples may enlarge known-class acceptance regions, increase
inter-device confusion, reduce the classification accuracy of authorized
devices, and make unknown devices more likely to be accepted as authorized
devices.

This work studies open-set WiFi RFF by combining DeepCRF with OpenMax.
DeepCRF uses channel simulation and Gaussian noise augmentation to learn device representations across different channel
conditions. After model training, OpenMax uses the activation vectors of
correctly classified training samples to calculate a mean activation vector
for each known class. It then calculates the distance from each activation
vector to the corresponding mean activation vector and selects the samples
with the largest distances to fit a Weibull distribution. During inference,
the fitted distribution is used to determine whether a test sample remains
within the acceptance region of a known class.

However, correct classification alone does not ensure that a sample is
suitable for OpenMax construction. Some augmented samples may be classified
correctly while their activation vectors deviate from the typical activation
pattern of their source devices and exhibit low classification confidence.
Including these samples may shift the mean activation vector away from the
typical class representation and cause large deviations to be treated as
normal variations of known classes. As a result, known-class acceptance
regions may become larger, increasing confusion among authorized devices and
the risk of accepting unknown devices as authorized devices.

To address these issues, we propose C\textsuperscript{2}T-OpenMax, which
contains two components: center-constrained learning and
confidence-guided tail modeling. We introduce center-constrained learning
during model training to constrain the features produced by the feature
extractor and improve compactness within each known class.
Confidence-guided tail modeling selects only correctly classified,
high-confidence training samples to construct more representative mean
activation vectors and Weibull tail distributions. Together, the two
components improve authorized device classification and unknown device
detection while preserving the environmental generalization provided by
data augmentation. Fig.~\ref{fig:motivation} illustrates the studied problem
and the corresponding solutions. Our main contributions are summarized as follows:
\begin{itemize}
	\item We analyze the difficulty of applying OpenMax to an augmented WiFi
	RFF model and show that correctly classified samples with low
	classification confidence may lead to unreliable known-class modeling.
\vspace{-0.1cm}
	\item We introduce center-constrained learning during model training to
	constrain the features produced by the feature extractor and improve
	compactness within each known class.
\vspace{-0.1cm}
	\item We propose confidence-guided tail modeling, which uses only
	correctly classified, high-confidence training samples to construct
	more representative mean activation vectors and Weibull tail
	distributions.
\end{itemize}

\section{Related Work}

\subsection{Deep Learning and Channel-Robust RF Fingerprinting}

Deep learning has advanced RFF from hand-crafted signal features toward
end-to-end device representation learning~\cite{riyaz2018deep}. For WiFi systems, CSI provides fine-grained
amplitude and phase responses that have been used for physical-layer device
identification~\cite{kong2023physical,kong2024csi,kong2024deepcrf,huang2025enhancing}. However, received CSI contains both
transmitter-dependent characteristics and environment-dependent channel
effects, making RFF models sensitive to changes in location, multipath
propagation, noise, and acquisition conditions.

Existing channel-robust approaches suppress such environmental variations
through feature purification, data augmentation, auxiliary objectives, or
contrastive learning~\cite{shen2022towards,zhang2023transmitter,kong2024deepcrf}. DeepCRF~\cite{kong2024deepcrf}, for example, combines
channel simulation, Gaussian-noise augmentation, and supervised contrastive
pretraining to improve cross-environment closed-set recognition. We adopt this
established closed-set framework and focus on its reliable extension to
open-set recognition, rather than proposing a new augmentation or
representation-learning method.

\subsection{Open-Set Recognition for RF Fingerprinting}

OSR requires a model to classify known classes while rejecting previously
unseen ones. Existing approaches mainly rely on confidence
scoring~\cite{hanna2020deep}, distribution modeling~\cite{bendale2016towards}, reconstruction~\cite{huang2022class}, or discriminative boundary
learning~\cite{chen2021adversarial,zhou2025robust}. Confidence-based methods
such as maximum softmax probability are simple to apply but can remain
overconfident on unknown samples. OpenMax~\cite{bendale2016towards} models the
upper tails of class-wise logit distances with Weibull distributions.
Reconstruction-based methods identify unknowns according to their mismatch
with known-class reconstruction patterns, whereas boundary-based methods learn
representations that separate known classes from open space.

Existing RFF-oriented OSR studies primarily improve unknownness scores,
decision boundaries, nuisance-feature suppression, or inference-time
adaptation~\cite{yang2025openrfi,li2026rff}. In contrast, we examine the reliability of the correctly
classified training activation vectors used to construct a known-class statistical
model. This issue is particularly relevant to OpenMax because both its class
mean activation vectors (MAVs) and Weibull tails are estimated from such
training samples.

\subsection{Known-Class Modeling under Augmented Representations}

Data augmentation is generally beneficial for RFF because it exposes a
closed-set model to broader channel variations and improves cross-environment
generalization. However, correctly classified samples in an augmented
representation may differ in confidence and in how well they represent the
core structure of a known class. These differences may have limited influence
on a closed-set decision but become important when the same activation vectors are
used to estimate class MAVs and upper-tail distance distributions.

Prior channel-robust RFF studies mainly evaluate representation robustness and
closed-set accuracy, while existing OSR methods mainly optimize unknown
rejection. The reliability of the known samples used for tail-distribution
modeling at the intersection of these two settings remains less explored. Our
work addresses this gap through representation-level compactness optimization
and confidence-guided refinement of the training samples used for OpenMax
construction.

\section{Problem Formulation and Motivation}

\subsection{Open-Set WiFi RFF Setting}

Let $\mathcal{Y}_{K}=\{1,\ldots,K\}$ denote the set of known-device
labels available during training. At test time, a CSI sample may come from
\[
\mathcal{Y}_{\mathrm{test}}
=
\mathcal{Y}_{K}\cup\mathcal{Y}_{U},
\]
where $\mathcal{Y}_{K}\cap\mathcal{Y}_{U}=\varnothing$ and
$\mathcal{Y}_{U}$ denotes previously unseen devices. Given an input sample $x$, the closed-set classifier produces
	a pre-softmax activation vector $\mathbf{v}(x)$ and a
	probability vector
	$\mathbf{p}(x)
	=\operatorname{softmax}(\mathbf{v}(x))$
over $\mathcal{Y}_K$, where $p(k\mid x)$ denotes the
probability assigned to class $k$. The open-set objective is to assign the correct device
label when $x$ belongs to a known class and to output \emph{unknown}
otherwise. Samples from $\mathcal{Y}_{U}$ are held out from MAV estimation,
Weibull fitting, and model construction.

\subsection{Tail-Modeling Challenge}

OpenMax constructs a statistical model for each known class from correctly
classified training logits. For class $k$, let
\[
\mathcal{C}_{k}
=
\left\{
x \mid y(x)=k,\ \hat{y}(x)=k
\right\}
\]
denote the set of correctly classified training samples, where
$\hat{y}(x)=\arg\max_j p(j\mid x)$. Conventional OpenMax computes the class mean
logit vector as
\[
\boldsymbol{\mu}_{\mathrm{base}}(k)
=
\frac{1}{|\mathcal{C}_{k}|}
	\sum_{x\in\mathcal{C}_{k}} \mathbf{v}(x),
\]
and fits a Weibull distribution to the upper tail of the distances from
$\mathbf{v}(x)$ to $\boldsymbol{\mu}_{\mathrm{base}}(k)$.

This construction implicitly treats all samples in $\mathcal{C}_{k}$ as
equally reliable. However, correct classification only indicates that the
ground-truth class receives the largest prediction score. A correctly
classified sample may still have low confidence, lie close to a decision
boundary, or be weakly representative of the class core. Such samples can
affect OpenMax construction in two related ways. First, including them in
the class average may shift the MAV away from the more representative
high-confidence region. Second, because all logit distances are defined
relative to the estimated MAV, this shift can change the distance values and
the ordering of samples selected for upper-tail fitting.

We therefore distinguish the complete correctly classified set
$\mathcal{C}_{k}$ from the confidence-screened subset
\[
\mathcal{S}_{k}
=
\left\{
x\in\mathcal{C}_{k}
\mid p(k\mid x)>\delta
\right\},
\qquad
\mathcal{S}_{k}\subseteq\mathcal{C}_{k},
\]
where $\delta$ is a confidence threshold. The purpose of screening is not to
question the benefit of augmentation or to remove samples from closed-set
training. Instead, it refines the observations used to summarize an augmented
known-class representation during offline OpenMax construction.

\subsection{Design Objectives}

A reliable open-set extension should satisfy two complementary objectives.
First, the learned known-device representations should be sufficiently compact
so that a class reference point captures nominal within-class behavior rather
than a highly dispersed collection of channel realizations. Second, the finite
sample set used to estimate this reference point and its upper-tail distance
distribution should contain reliable class representatives.

These objectives require different mechanisms. Representation-level
regularization changes the learned feature geometry, whereas confidence
screening changes only the subset used for MAV estimation and Weibull fitting.
Accordingly, we introduce center-constrained learning during the second
closed-set training stage and confidence-guided sample refinement during
offline OpenMax construction. The fitting-sample threshold is used only during
offline OpenMax construction, whereas inference combines OpenMax recalibrated
scores with a separately selected rejection threshold.

\section{Method}

\subsection{Overall Framework}

The adopted DeepCRF model uses two training stages. In Stage I, the
feature extractor is trained with supervised contrastive learning. In
Stage II, the feature extractor and classifier are optimized jointly.
C\textsuperscript{2}T-OpenMax keeps the first stage unchanged and adds center loss in the
second stage. After training, confidence screening is applied during
offline OpenMax construction.

\begin{figure}[t]
	\centering
	\par\smallskip
	\includegraphics[width=\linewidth]{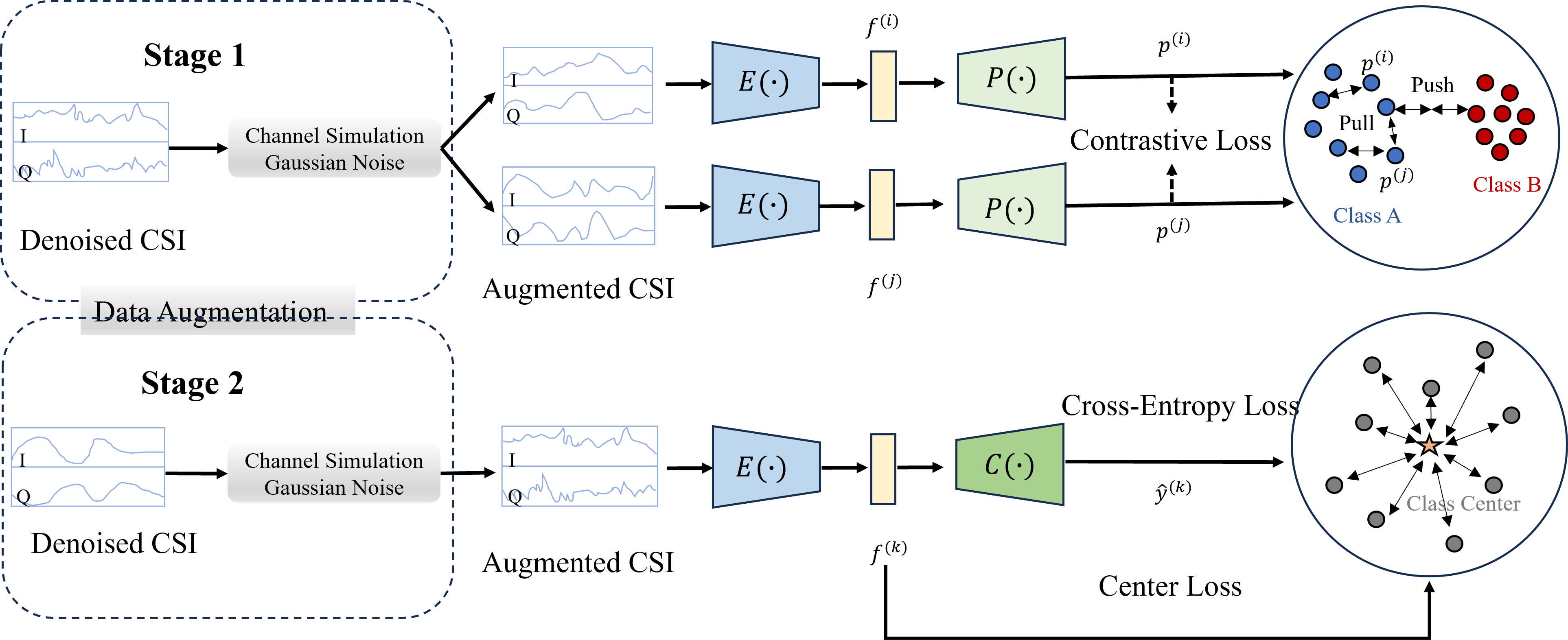}
	\par\smallskip
	\caption{Training framework of \method{}. The augmentation strategy and
		Stage I supervised contrastive learning are adopted from DeepCRF, while
		center loss is introduced in Stage II to improve intra-class
		compactness.}
	\label{fig:training-framework}
\end{figure}

\method{} extends the adopted augmented closed-set classifier with two
complementary components. Fig.~\ref{fig:training-framework} presents the
representation-learning stage. Channel-oriented augmentation and Stage I
supervised contrastive learning are adopted from DeepCRF, while center loss is
introduced in Stage II to improve intra-class compactness.
Fig.~\ref{fig:tail-modeling} presents the subsequent offline
OpenMax-construction stage, where confidence screening refines the training
subset used for MAV estimation and Weibull fitting. Algorithm~\ref{alg:ct_openmax}
summarizes the complete construction procedure.

\subsection{Adopted Augmented Closed-Set Framework}

Following DeepCRF~\cite{kong2024deepcrf}, the training set contains original
CSI samples and samples generated by channel simulation and Gaussian-noise
injection. In Stage I, only the feature extractor is optimized using supervised
contrastive learning (SupCon). For representation $\mathbf{f}_i$, the objective is
\begin{equation}
	\mathcal{L}_{\mathrm{supcon}}
	=
	\sum_{i=1}^{N}
	\frac{-1}{|P(i)|}
	\sum_{q\in P(i)}
	\log
	\frac{
		\exp\left(\mathbf{f}_i^{\top}\mathbf{f}_q/\tau\right)
	}{
		\sum_{a\in A(i)}
		\exp\left(\mathbf{f}_i^{\top}\mathbf{f}_a/\tau\right)
	}.
\end{equation}
where $N$ is the number of samples in the contrastive training batch,
$P(i)$ contains the indices of samples sharing the label of sample $i$,
$A(i)$ contains all sample indices except $i$, and $\tau$ is the
temperature parameter. Both the
augmentation procedure and Stage I training are adopted components rather than
contributions of this work.

\subsection{Center-Constrained Feature Learning}

In Stage II, the feature extractor and closed-set classifier are optimized
jointly. Let $\mathcal{B}$ denote a training mini-batch with
$M=|\mathcal{B}|$ samples. Let $p(k\mid x)$ denote the predicted probability of class $k$ for sample
$x$, and let $y_k(x)$ denote its one-hot label. The cross-entropy loss is
\begin{equation}
	\mathcal{L}_{\mathrm{ce}}
	=
	-\frac{1}{M}
	\sum_{x\in\mathcal{B}}
	\sum_{k=1}^{K}
	y_k(x)\log p(k\mid x).
\end{equation}

For learned feature $\mathbf{f}(x)$ and the center $\mathbf{c}(y(x))$ of its ground-truth class,
the center loss is
\begin{equation}
	\mathcal{L}_{\mathrm{center}}
	=
	\frac{1}{2M}
	\sum_{x\in\mathcal{B}}
		\left\lVert \mathbf{f}(x)-\mathbf{c}(y(x))\right\rVert_{2}^{2}.
\end{equation}

Although center loss is applied to the intermediate features, its effect is
propagated to the classifier outputs used by OpenMax. By reducing the
within-class variation of the input representations to the classifier, it
also promotes more compact class-wise logit distributions, providing a more
stable basis for MAV estimation and distance-based tail modeling.

The Stage II objective is
\begin{equation}
	\mathcal{L}_{\mathrm{hybrid}}
	=
	\mathcal{L}_{\mathrm{ce}}
	+
	\lambda\mathcal{L}_{\mathrm{center}},
\end{equation}
where $\lambda$ balances classification and compactness objectives. Center loss
directly reduces intra-class variation and provides a more compact
known-device representation for subsequent statistical modeling.
Inter-class discrimination is learned jointly through supervised contrastive
learning and cross-entropy optimization.

\begin{figure}[ht]
	\centering
	\par\smallskip
	\includegraphics[width=\linewidth,trim=0 8 0 0,clip]{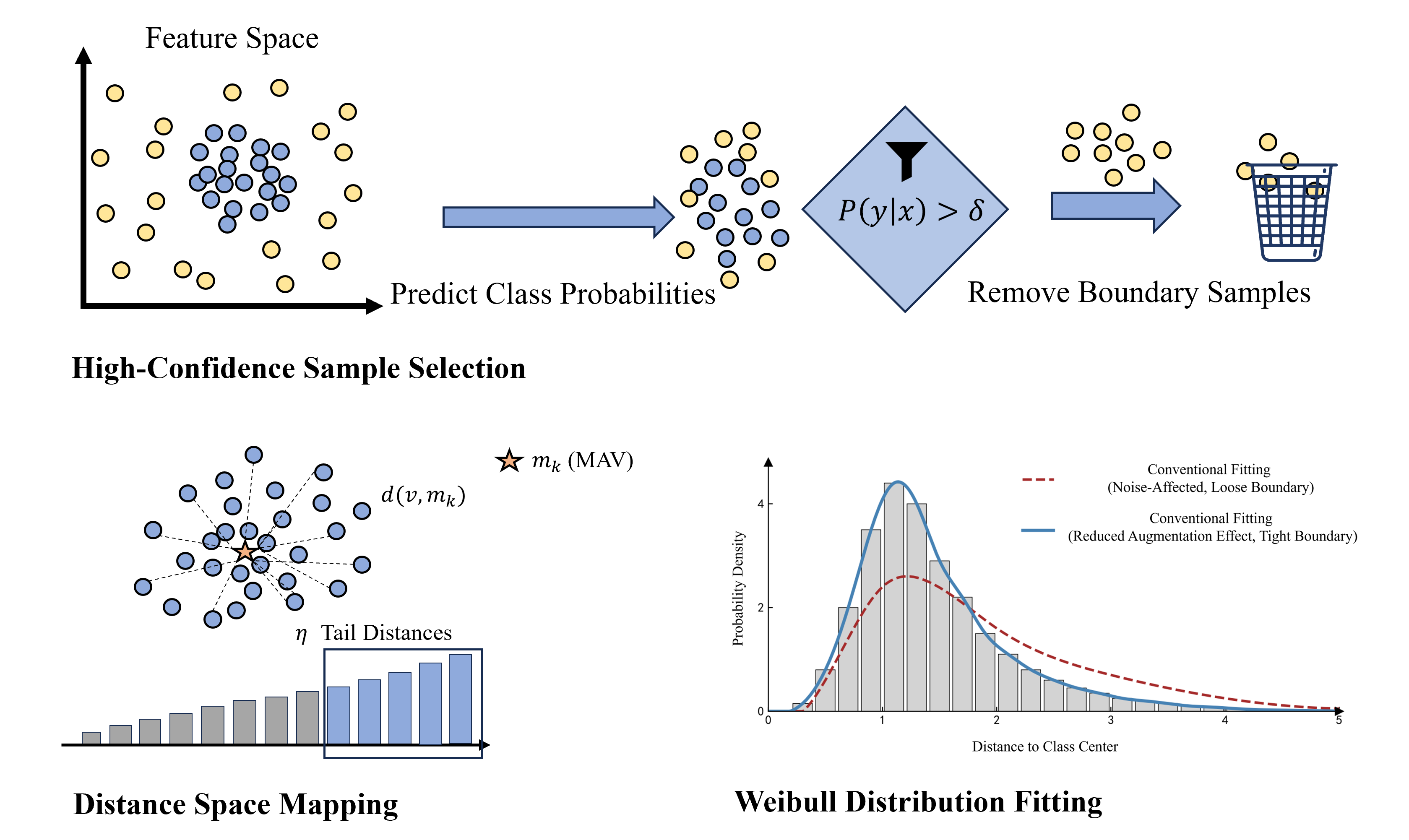}
	\par\smallskip
	\caption{Confidence-guided OpenMax construction. Correctly classified training
		samples are further screened according to the predicted probability assigned
		to their ground-truth class. The retained training logits are used for MAV estimation
		and Weibull tail fitting.}
	\label{fig:tail-modeling}
\end{figure}

\subsection{Confidence-Guided Tail Modeling}

After closed-set training, the class-wise OpenMax models are constructed
offline. Let $\mathbf{f}(x)=F(x)$ denote the intermediate feature representation,
and let $\mathbf{v}(x)=G(\mathbf{f}(x))$ denote the pre-softmax activation
vector produced by the classifier. The corresponding probability vector is
$\mathbf{p}(x)=\operatorname{softmax}(\mathbf{v}(x))$, where
$p(k\mid x)$ denotes the probability assigned to class $k$.
The predicted class is
$\hat{y}(x)=\arg\max_j p(j\mid x)$. The screened set
for class $k$ is
\begin{equation}
	\mathcal{S}_k
	=
	\left\{
	x\in\mathcal{C}_k
	\mid p(k\mid x)>\delta
	\right\},
	\qquad
	\mathcal{S}_k\subseteq\mathcal{C}_k.
	\label{eq:screen}
\end{equation}
Here, $p(k\mid x)$ is the predicted probability assigned to the ground-truth
class. Screening changes only the observations used for offline OpenMax
construction; it neither modifies the trained feature extractor nor removes
samples from closed-set training.

The screened MAV is computed as
\begin{equation}
	\boldsymbol{\mu}(k)
	=
	\frac{1}{|\mathcal{S}_k|}
		\sum_{x\in\mathcal{S}_k}\mathbf{v}(x).
	\label{eq:mav}
\end{equation}
For each retained logit vector, we calculate a mixed Euclidean--cosine distance:
\begin{equation}
	d(x,k)
	=
		\frac{\left\lVert \mathbf{v}(x)-\boldsymbol{\mu}(k)\right\rVert_2}{200}
	+
	\left(
	1-
		\frac{\mathbf{v}(x)^{\top}\boldsymbol{\mu}(k)}
		{\left\lVert \mathbf{v}(x)\right\rVert_2
			\left\lVert \boldsymbol{\mu}(k)\right\rVert_2}
	\right).
	\label{eq:distance}
\end{equation}

For class $k$, the distances
$\mathcal{D}_k=\{d(x,k)\mid x\in\mathcal{S}_k\}$ are sorted in descending order.
The largest proportion $r$ is retained as the tail set
$\mathcal{T}_k=
\operatorname{Top}_{\lceil r|\mathcal{D}_k|\rceil}(\mathcal{D}_k)$, and the
class-specific Weibull model is
$\mathcal{W}_k=\operatorname{WeibullFit}(\mathcal{T}_k)$. In
our implementation, $\delta=0.8$ and $r=0.05$. The same screened subset is
therefore used to determine the class reference point and the distance
distribution from which its upper tail is selected.

\begin{algorithm}[t]
		\caption{C\textsuperscript{2}T-OpenMax Construction}
		\label{alg:ct_openmax}
	\small
	\begin{algorithmic}[1]
		\Require Known-device training set
		$\mathcal{D}=\{(x,y(x))\}$;
		trained feature extractor $F$ and classifier $G$;
		confidence threshold $\delta$; tail ratio $r$
		\Ensure Class MAVs $\{\boldsymbol{\mu}(k)\}_{k=1}^{K}$ and Weibull models
		$\{\mathcal{W}_k\}_{k=1}^{K}$
		
		\For{$\text{each }x\in\mathcal{D}$}
			\State $\mathbf{f}(x) \leftarrow F(x)$
			\State $\mathbf{v}(x) \leftarrow G(\mathbf{f}(x))$
			\State $\mathbf{p}(x) \leftarrow
			\operatorname{softmax}(\mathbf{v}(x))$
			\State $\hat{y}(x) \leftarrow \arg\max_j p(j\mid x)$
		\EndFor
		
		\For{$k=1$ to $K$}
			\State $\mathcal{S}_k \leftarrow
			\{x\mid y(x)=k,\hat{y}(x)=k,p(k\mid x)>\delta\}$
			\State $\boldsymbol{\mu}(k) \leftarrow
			|\mathcal{S}_k|^{-1}\sum_{x\in\mathcal{S}_k}\mathbf{v}(x)$
			\State $\mathcal{D}_k \leftarrow
			\{d(x,k)\mid x\in\mathcal{S}_k\}$
		\State Sort $\mathcal{D}_k$ in descending order
			\State $\mathcal{T}_k \leftarrow$ largest
		$\lceil r|\mathcal{D}_k|\rceil$ distances in $\mathcal{D}_k$
			\State $\mathcal{W}_k \leftarrow
		\operatorname{WeibullFit}(\mathcal{T}_k)$
		\EndFor
		
		\State \Return $\{\boldsymbol{\mu}(k),\mathcal{W}_k\}_{k=1}^{K}$
	\end{algorithmic}
\end{algorithm}

Fig.~\ref{fig:tail-modeling} illustrates this construction process. The
screening operation is applied before MAV estimation, and the resulting MAV is
then used to compute and rank the retained logit distances.

\subsection{Open-Set Inference}

For a test CSI sample $x$, the closed-set classifier produces a pre-softmax
logit vector and known-class scores. Following OpenMax~\cite{bendale2016towards},
each class-specific Weibull model produces an outlier weight from the distance
between the test logit vector and the corresponding MAV. The weight discounts
the associated known-class score, and the removed score mass is accumulated
into the unknown class.

Let $\widetilde{p}(c\mid x)$ denote the OpenMax recalibrated score for class
$c\in\mathcal{Y}_K\cup\{u\}$. Let
\begin{equation}
	k^{*}
	=
	\arg\max_{k\in\mathcal{Y}_{K}}
	\widetilde{p}(k\mid x)
\end{equation}
denote the known class with the largest recalibrated score. The final decision
is
\begin{equation}
		\hat{y}^{\mathrm{open}}(x)
	=
	\begin{cases}
		u,
		&
		\widetilde{p}(u\mid x)
		>
		\widetilde{p}(k^{*}\mid x)
		\ \text{or}\
		\widetilde{p}(k^{*}\mid x)<\gamma,
		\\[2mm]
		k^{*},
		&
		\text{otherwise},
	\end{cases}
	\label{eq:open-set-decision}
\end{equation}
where $u$ denotes the unknown class. Here, $\delta$ is the fitting-sample
screening threshold used only during offline OpenMax construction, whereas
$\gamma$ is the inference-time rejection threshold. The threshold $\gamma$ is
selected on the known-device validation set according to classification
accuracy, without using any unknown-device samples.

\section{Experiments}

\subsection{Experimental Setup}

\paragraph{Dataset and protocols.}
We use the public WiFi CSI dataset released with
DeepCRF~\cite{DeepCRF2025}. The dataset contains CSI measurements collected
from multiple WiFi devices at nine locations over approximately one year. It
covers diverse propagation and acquisition conditions, including line-of-sight
(LOS) and non-line-of-sight (NLOS) links, static and dynamic scenarios, and
indoor and outdoor environments. We follow the data partition and augmentation
protocol of DeepCRF~\cite{kong2024deepcrf}. P1 and P2 are used for model
development, and their held-out test samples are jointly reported as P1--P2.
P3--P9 are used for cross-location testing. 

The cross-location experiment contains 15 known devices and 4 unknown devices.
The openness experiment fixes 12 known devices and increases the number of
unknown devices from 1 to 7. The ablation study uses 12 known and 7 unknown
devices, corresponding to the largest openness setting.

\paragraph{Baselines and metrics.}
We compare \method{} with maximum softmax probability
(MSP)~\cite{hanna2020deep}, original OpenMax~\cite{bendale2016towards},
class-specific semantic reconstruction (CSSR)~\cite{huang2022class}, and
adversarial reciprocal points learning (ARPL)~\cite{chen2021adversarial}.
These methods represent confidence-based, distribution-based,
reconstruction-based, and discriminative-boundary OSR strategies,
respectively. All methods use identical known/unknown device partitions and
the same augmented training data; in particular, CSSR and ARPL use the same
augmentation protocol.

We report open-set accuracy, the area under the receiver operating
characteristic curve (AUROC), and the open-set classification rate (OSCR).
Open-set accuracy jointly
considers correct known-device classification and correct unknown-device
rejection:
\begin{equation}
	\mathrm{Accuracy}
	=
	\frac{
		N_{\mathrm{known\mbox{-}correct}}
		+
		N_{\mathrm{unknown\mbox{-}correct}}
	}{
		N_{\mathrm{total}}
	}.
\end{equation}
AUROC measures known--unknown separability. OSCR additionally considers whether an accepted known sample is
assigned to its correct device class and therefore evaluates the trade-off
between correct known-device classification and false acceptance of unknown
devices~\cite{scheirer2012toward}.

\begin{table*}[ht]
	\centering
	\label{tab:locations}
	\small
	\begin{tabular}{lcccccccc}
		\toprule
		Method & P1--P2 & P3 & P4 & P5 & P6 & P7 & P8 & P9 \\
		\midrule
		MSP
		& 85.83 & 83.21 & 79.24 & 69.99
		& 85.48 & 70.87 & 75.95 & 70.41 \\
		
		OpenMax
		& 85.90 & 82.82 & 79.64 & 69.18
		& 84.21 & 71.33 & 75.12 & 69.56 \\
		
		CSSR
		& 86.12 & 84.88 & \textbf{86.37} & 78.92
		& 88.90 & 75.57 & 81.67 & 70.34 \\
		
		ARPL
		& 87.97 & 80.13 & 84.32 & 89.77
		& 94.05 & 82.78 & 92.85 & 66.37 \\
		
		\method
		& \textbf{95.70} & \textbf{86.14} & 80.46
		& \textbf{90.26} & \textbf{96.35} & \textbf{87.38}
		& \textbf{93.46} & \textbf{84.61} \\
		\bottomrule
	\end{tabular}
	\caption{Open-set accuracy (\%) across different locations. Bold indicates
		the best result in each column. P1 and P2 are jointly reported.}
\end{table*}

\paragraph{Implementation details.}
The closed-set training configuration, including data augmentation and
two-stage optimization, follows DeepCRF~\cite{kong2024deepcrf}. Center loss is
added only in Stage II with $\lambda=0.005$, while Stage I remains unchanged.
After training, the pre-softmax logits of correctly classified samples are used
for OpenMax construction. We set $\delta=0.8$ and use the largest $5\%$ of
class-wise distances for Weibull fitting. The mixed Euclidean--cosine distance
follows Eq.~\eqref{eq:distance} and is used in both construction and inference.
The inference threshold $\gamma$ is selected using only known-device validation
samples and fixed during testing. No unknown-device sample is used for MAV
estimation, Weibull fitting, or threshold selection, and the same
known/unknown assignment is used across locations.

\subsection{Comparison Across Locations}

Table~\ref{tab:locations} reports the open-set accuracy under the
15-known and 4-unknown protocol. \method{} achieves the best result at seven
of the eight reported location groups and consistently outperforms original
OpenMax across all locations. At P1--P2, the accuracy increases from 85.90\%
to 95.70\%, showing that the proposed method is effective not only under
cross-location shifts but also on held-out real samples collected at the
locations involved in model development.

The improvements become more pronounced in several cross-location settings.
In particular, the gains over original OpenMax reach 21.08 and 18.34
percentage points at P5 and P8, respectively. Consistent improvements are also
observed at P6, P7, and P9, indicating that compact representation learning
and confidence-guided fitting-sample refinement provide more reliable
class-wise statistics under substantial channel variations.
P4 is characterized by severe multipath propagation~\cite{kong2024deepcrf}.
Although \method{} improves original OpenMax from 79.64\% to 80.46\%, it
remains below CSSR and ARPL at this location. This result suggests that severe
multipath can distort the learned representation and classifier logits in a
manner that cannot be fully addressed by feature compactness and OpenMax
fitting-sample refinement alone.

\begin{figure}[ht]
	\centering
	\includegraphics[width=\linewidth]{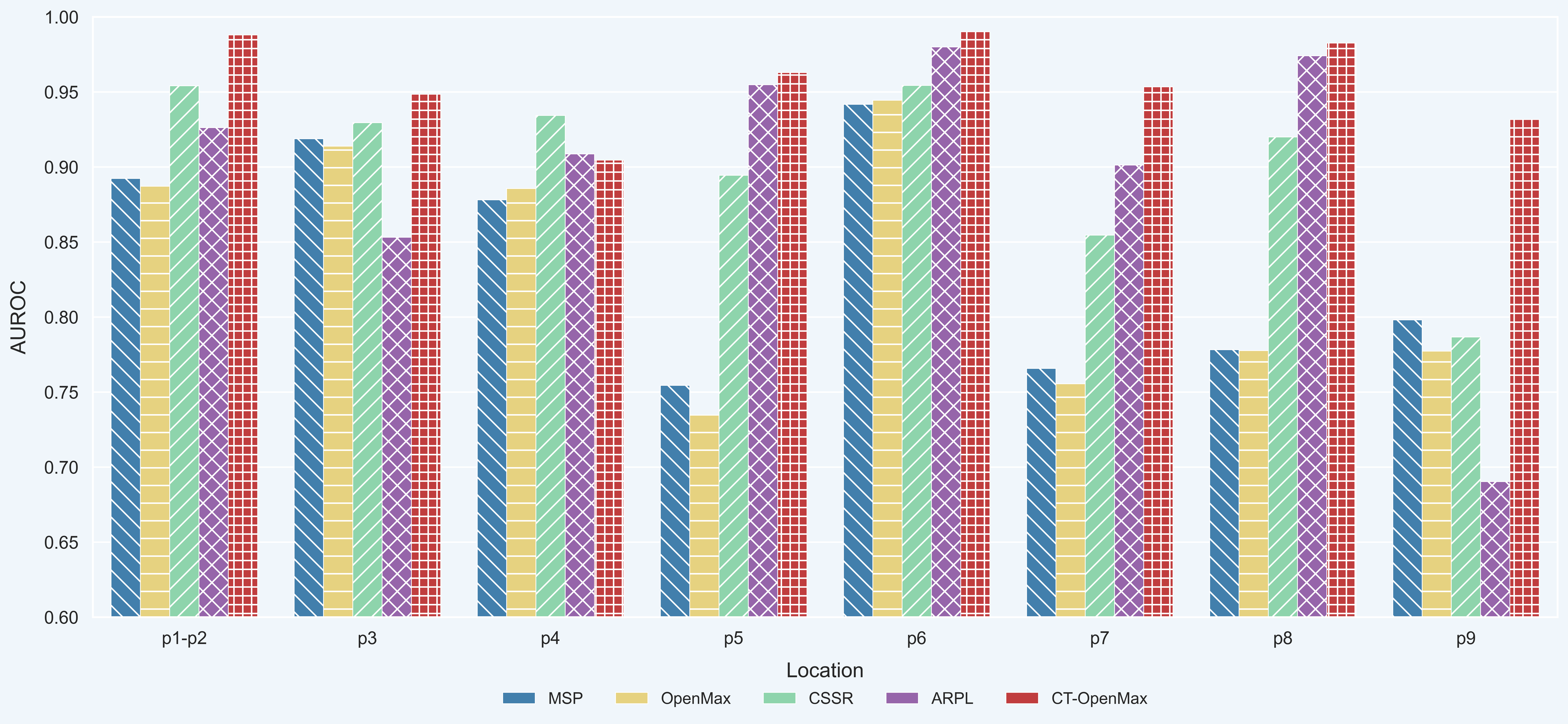}
	\caption{AUROC across different locations.}
	\label{fig:location-auroc}
\end{figure}

\begin{figure}[ht]
	\centering
	\includegraphics[width=\linewidth]{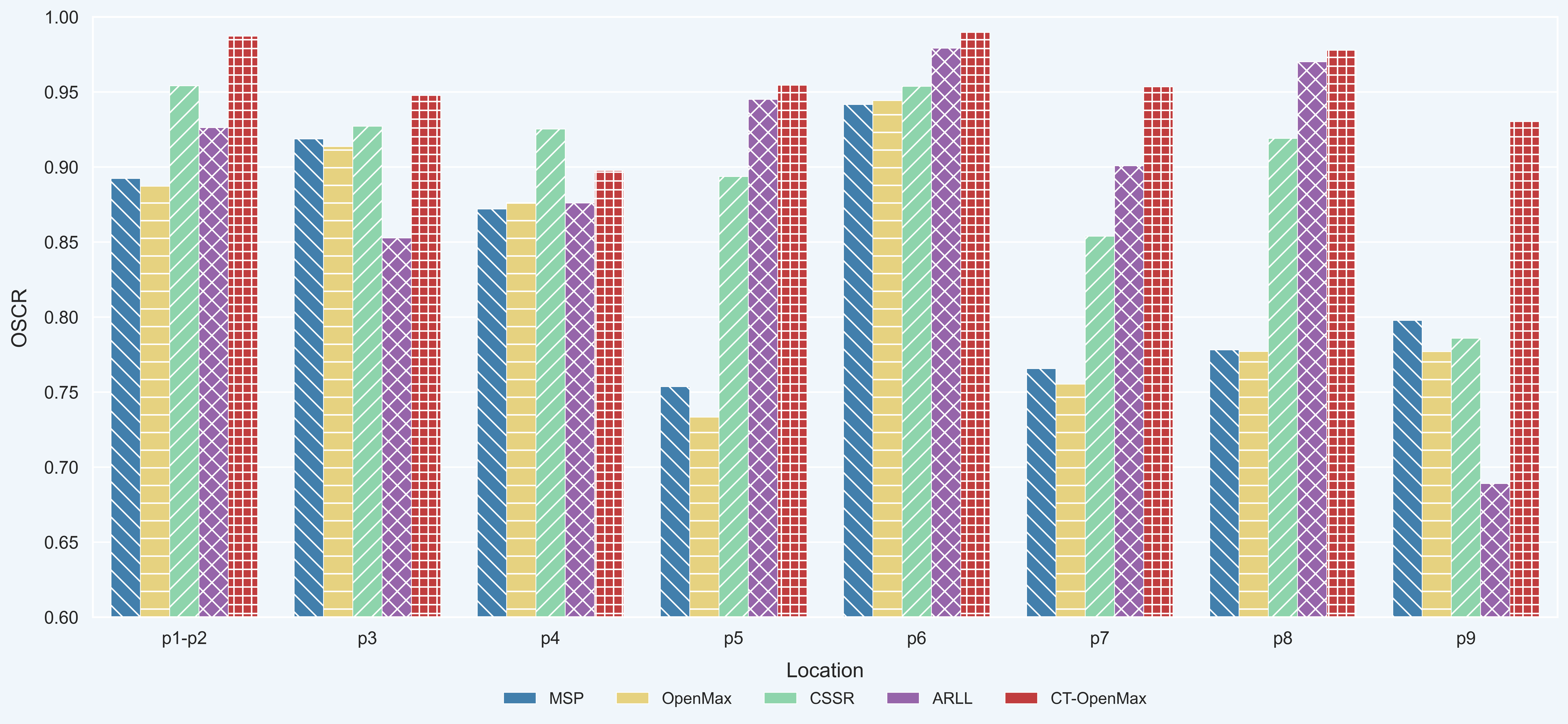}
	\caption{OSCR across different locations.}
	\label{fig:location-oscr}
\end{figure}

Fig.~\ref{fig:location-auroc} and Fig.~\ref{fig:location-oscr} present the
corresponding AUROC and OSCR results. Their overall trends are consistent with
the accuracy comparison, indicating that the improvements are reflected in
both known--unknown separability and correct known-device recognition rather
than being caused only by a particular decision threshold.


\subsection{Performance Under Different Openness}

We fix 12 known devices and gradually increase the number of unknown devices
from 1 to 7. Following the openness definition of Scheirer et al.\cite{scheirer2012toward}, the corresponding openness ranges from 2.02\% to
12.01\%. For each openness setting, test samples from all locations are
combined before AUROC and OSCR are calculated.

\begin{figure}[t]
	\centering
	\begin{subfigure}[t]{0.48\linewidth}
		\centering
		\includegraphics[width=\linewidth]{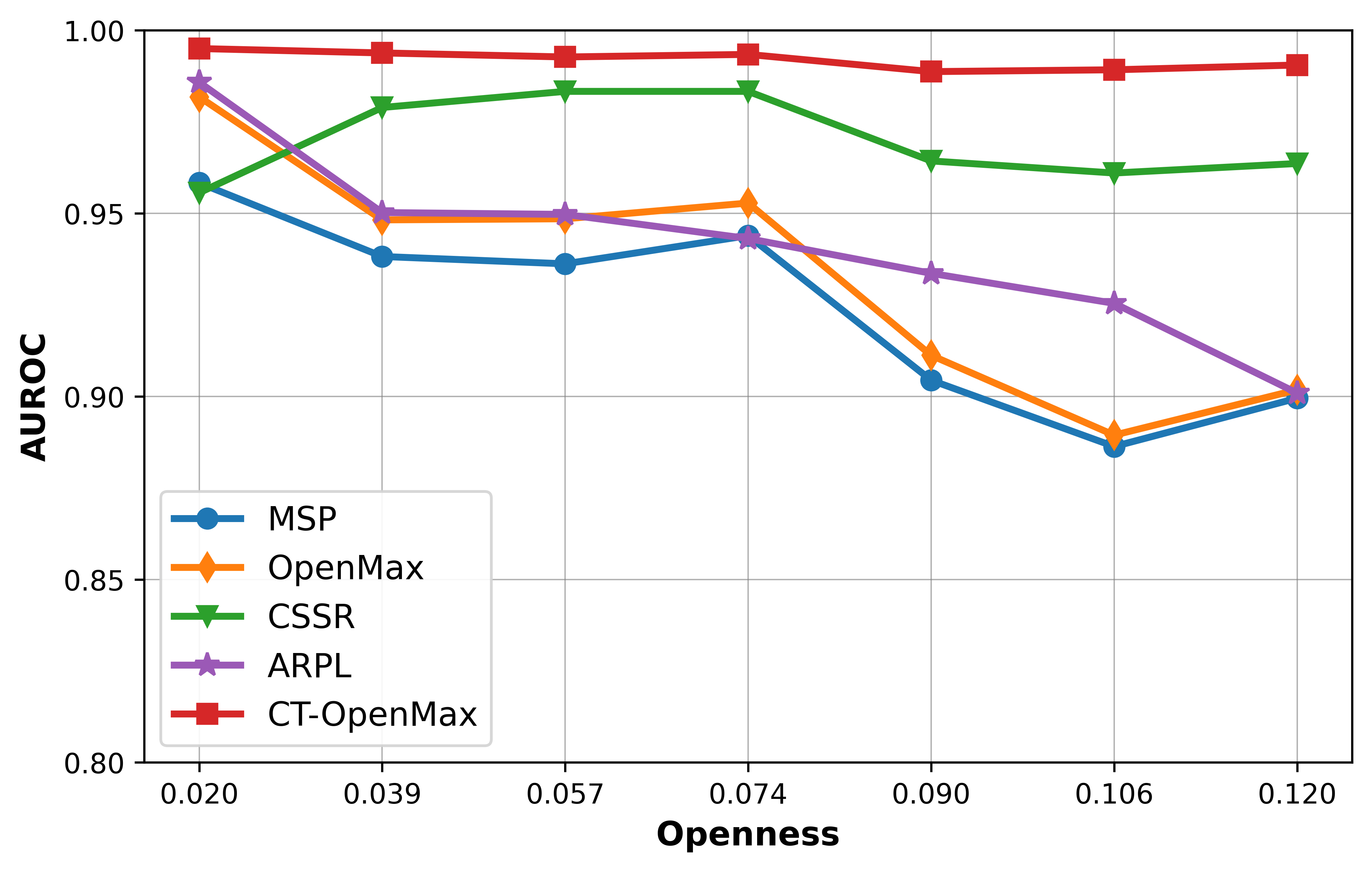}
		\caption{AUROC}
		\label{fig:openness-auroc}
	\end{subfigure}\hfill
	\begin{subfigure}[t]{0.48\linewidth}
		\centering
		\includegraphics[width=\linewidth]{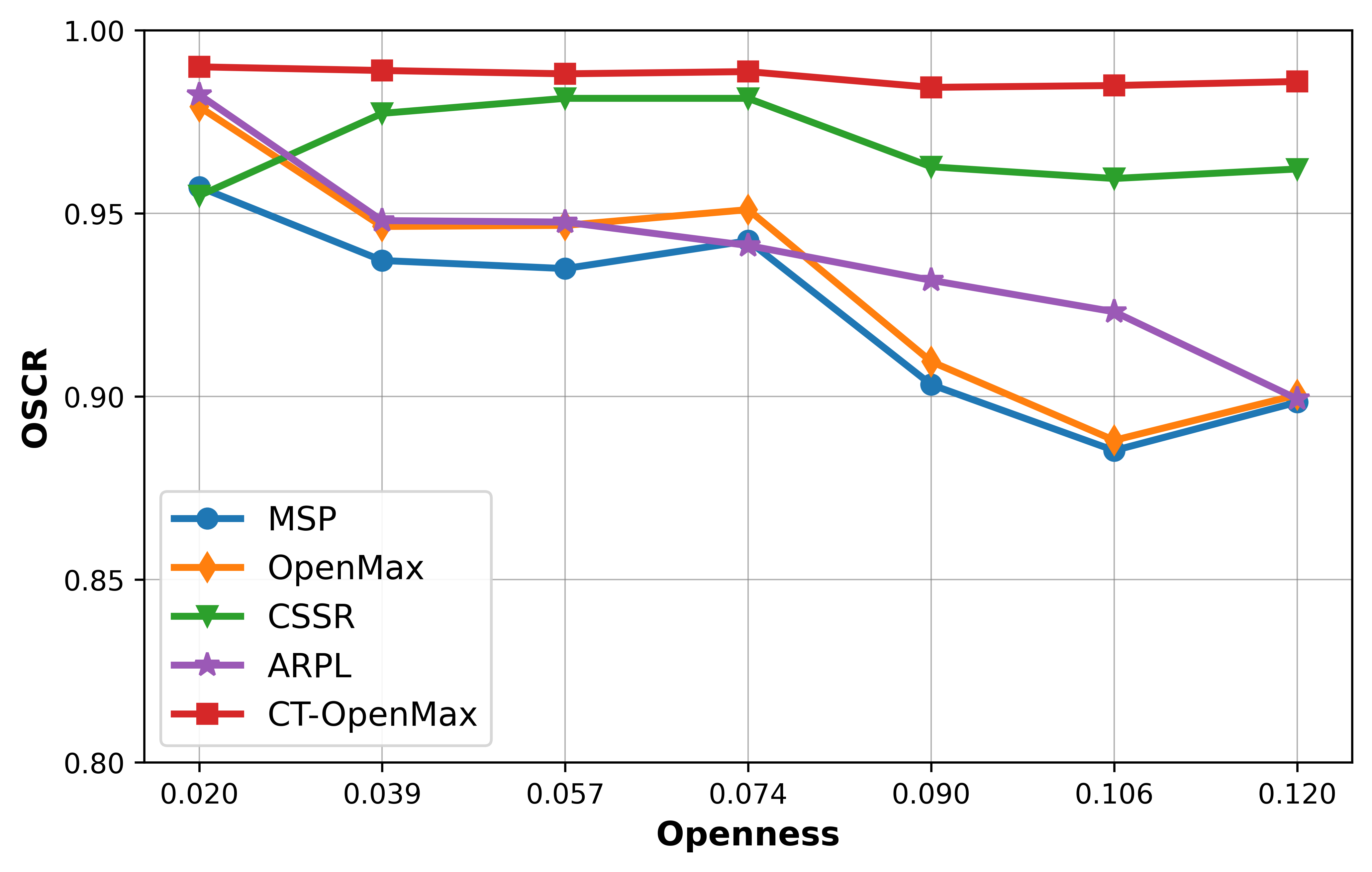}
		\caption{OSCR}
		\label{fig:openness-oscr}
	\end{subfigure}
	\caption{Open-set performance with 12 known devices and an increasing number
		of unknown devices. Test samples from all locations are combined for
		evaluation.}
	\label{fig:openness-results}
\end{figure}

As shown in Fig.~\ref{fig:openness-auroc}
and Fig.~\ref{fig:openness-oscr}, the performance of the evaluated methods
generally decreases or fluctuates as more unknown devices are introduced.
The variation is not strictly monotonic because different unknown devices can
have different degrees of similarity to the known-device classes. Therefore,
adding an unknown class does not necessarily increase the recognition
difficulty by the same amount at every openness level.

Across all reported openness settings, \method{} achieves higher AUROC and
OSCR than the compared methods. These results indicate that the method
maintains reliable known--unknown separability when the composition
of the unknown-device set changes.

AUROC and OSCR exhibit similar overall trends because both metrics depend on
the confidence ordering between known and unknown samples. However, OSCR is
more restrictive because a known sample contributes positively only when it
is both accepted and assigned to its correct known-device class. Consistent
performance on both metrics therefore indicates that \method{} improves unknown
rejection while preserving useful known-device identification.

\subsection{Ablation Study}

Table~\ref{tab:ablation} evaluates the effects of center-constrained learning
and confidence-guided sample screening. The experiment uses 12 known devices
and 7 unknown devices, corresponding to the largest openness setting. Test
samples from all locations are combined before Accuracy, AUROC, and OSCR are
calculated. The base model denotes the adopted augmented closed-set model
combined with original OpenMax.

Adding center-constrained learning improves accuracy from 83.60\% to 88.25\%,
corresponding to an increase of 4.65 percentage points. AUROC and OSCR
increase by 0.0516 and 0.0501, respectively. These results indicate that
improving intra-class compactness provides a more stable feature distribution
for subsequent OpenMax construction.

Confidence-guided sample screening alone increases accuracy by 7.13
percentage points, AUROC by 0.0733, and OSCR by 0.0740. The larger individual
gain obtained by this component indicates that correctly classified samples
are not equally reliable for MAV estimation and Weibull fitting. Refining the
fitting subset therefore substantially affects the resulting known-class
statistical models.

Combining both components produces the strongest performance. Compared with
the base model, \method{} improves accuracy by 12.31 percentage points, AUROC
by 0.0887, and OSCR by 0.0856. It also improves accuracy by 5.18 percentage
points over confidence screening alone. These results support the intended
complementarity of the two components: center loss changes the learned feature
geometry, whereas confidence screening changes the samples used to estimate
the class-wise statistical model.

\begin{table}[ht]
	\centering
	\small
	\begin{tabular}{lccc}
		\toprule
		Method & Accuracy (\%) & AUROC & OSCR \\
		\midrule
		Base & 83.60 & 0.9018 & 0.9004 \\
		Base + Center & 88.25 & 0.9534 & 0.9505 \\
		Base + Screening & 90.73 & 0.9751 & 0.9744 \\
		\method & \textbf{95.91} & \textbf{0.9905} & \textbf{0.9860} \\
		\bottomrule
	\end{tabular}

	\caption{Ablation results under the 12-known and 7-unknown setting. Test
		samples from all locations are combined for evaluation. Base denotes the
		adopted augmented closed-set model with original OpenMax.}
	\label{tab:ablation}
\end{table}

\subsection{Visualization of Training Samples Used for Weibull Fitting}

\begin{figure}[t]
	\centering
	\begin{subfigure}[t]{0.48\linewidth}
		\centering
		\includegraphics[width=\linewidth]{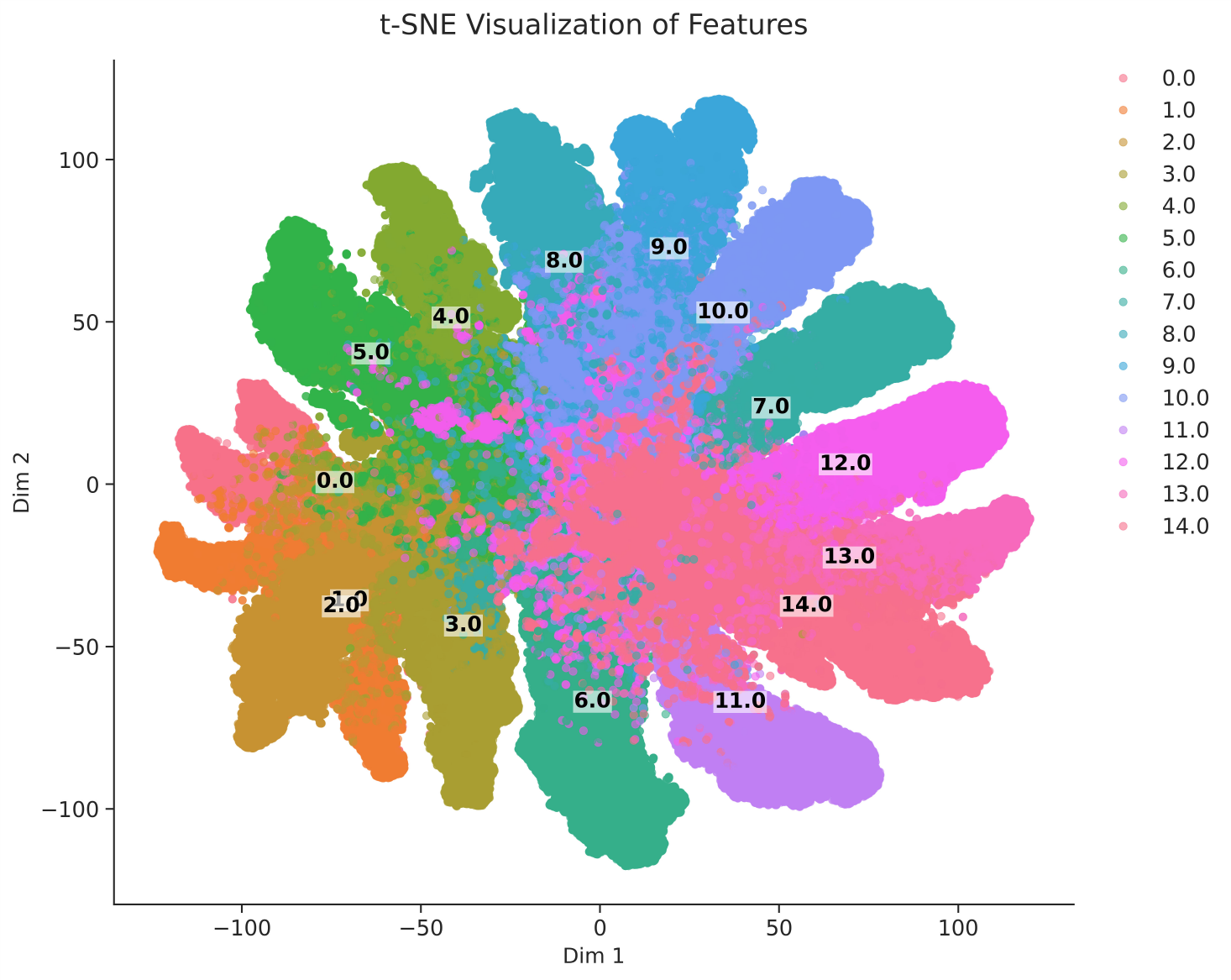}
		\caption{Original OpenMax}
	\end{subfigure}\hfill
	\begin{subfigure}[t]{0.48\linewidth}
		\centering
		\includegraphics[width=\linewidth]{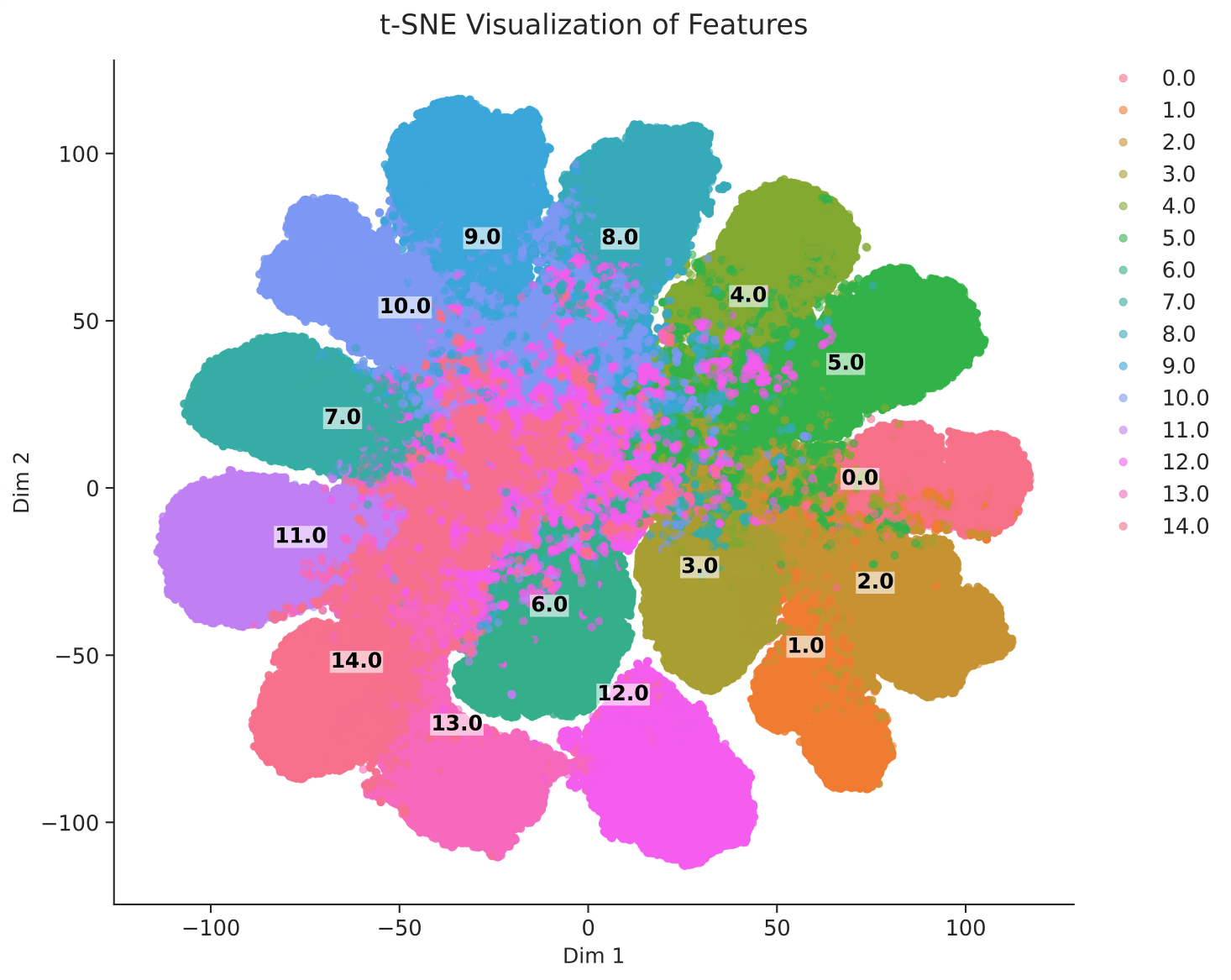}
		\caption{Base + Center}
	\end{subfigure}\par\smallskip
	\begin{subfigure}[t]{0.48\linewidth}
		\centering
		\includegraphics[width=\linewidth]{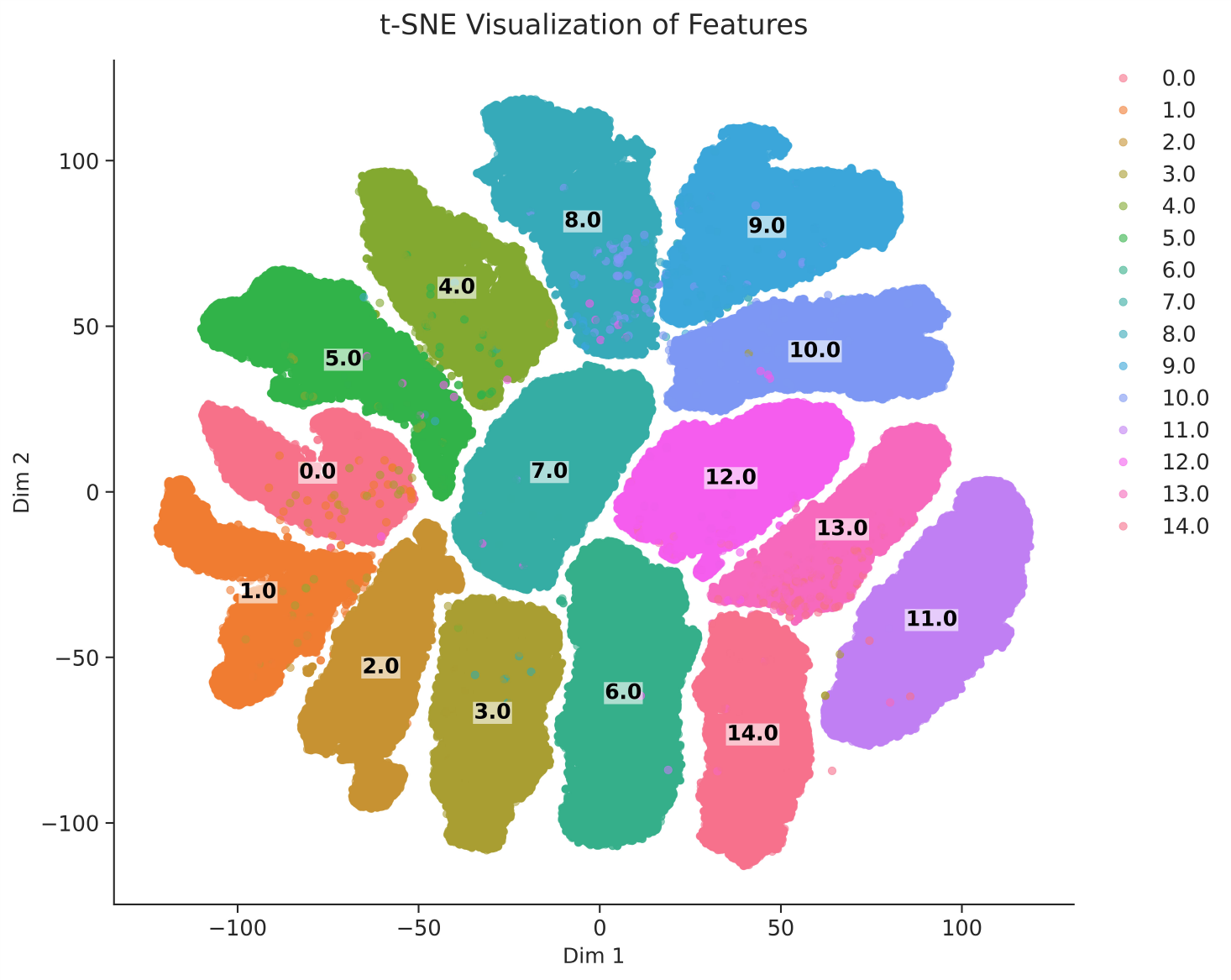}
			\caption{Base + Screening}
	\end{subfigure}\hfill
	\begin{subfigure}[t]{0.48\linewidth}
		\centering
		\includegraphics[width=\linewidth]{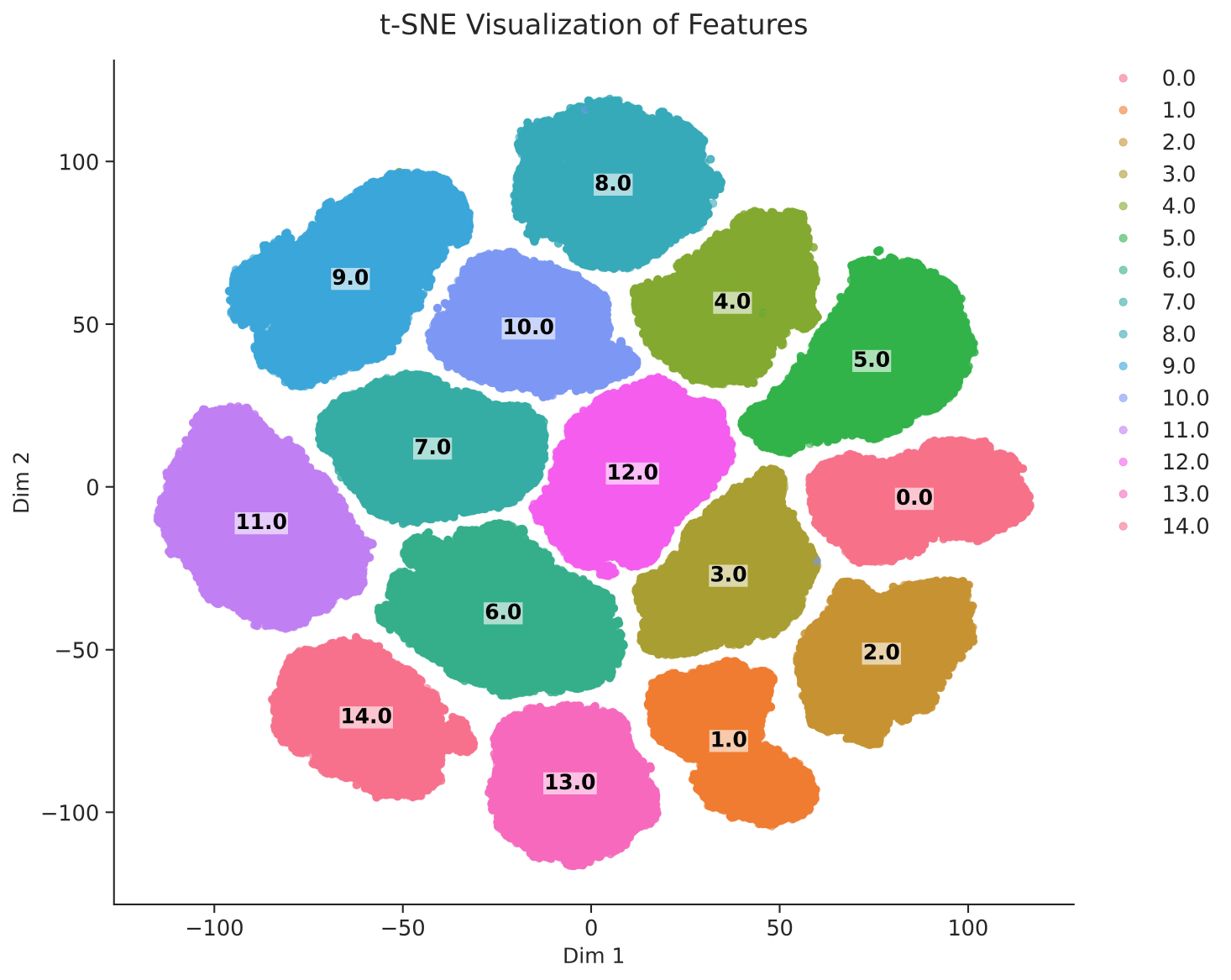}
		\caption{\method}
	\end{subfigure}
	\caption{t-SNE visualization of the class-wise training logits retained for
		MAV estimation and subsequent tail selection. Confidence screening modifies
		only the OpenMax construction subset and does not change the trained feature
		extractor.}
	\label{fig:tsne}
\end{figure}

OpenMax constructs its MAVs and Weibull models from training logits.
Fig.~\ref{fig:tsne} therefore visualizes the class-wise training logits used
for MAV estimation and Weibull fitting, rather than the complete test-logit
distribution.

Original OpenMax uses all correctly classified training samples. Introducing
center loss changes the learned feature representation and improves
intra-class compactness. Confidence screening, in contrast, leaves the
feature extractor unchanged and removes only correctly classified samples
that do not satisfy the confidence requirement.

The screening-only visualization should therefore be interpreted as a change
in the fitting subset rather than a change in the learned feature space. These
plots provide qualitative evidence of the distinct roles of the two components;
they do not indicate that confidence screening changes the network
representation.

\section{Conclusion}

This work studies how to reliably extend an augmented closed-set WiFi RFF
model to open-set recognition. \method{} combines center-constrained learning,
which improves intra-class compactness, with confidence-guided tail modeling,
which refines the training logits used for MAV estimation and Weibull tail
fitting. Experiments across locations, openness settings, ablations, and
fitting-sample visualizations demonstrate the complementary effects of the two
components. Although implemented with OpenMax, the sample-refinement strategy
may also benefit other distance- or tail-distribution-based OSR methods.
Severe multipath conditions remain challenging, and future work will explore
adaptive screening, more robust tail modeling, and broader validation across
devices, receivers, and channel environments.

%\section{Acknowledgments}

\bibliography{cite}
\end{document}